\documentstyle[emulateapj,graphicx]{article}
\def\gsim{\;\rlap{\lower 2.5pt
 \hbox{$\sim$}}\raise 1.5pt\hbox{$>$}\;}
\def\lsim{\;\rlap{\lower 2.5pt
   \hbox{$\sim$}}\raise 1.5pt\hbox{$<$}\;}

\newcommand{\beq}{\begin{equation}}
\newcommand{\eeq}{\end{equation}}
\def\myputfigure#1#2#3#4#5%
{\hskip0.03\textwidth\vskip#5pt%\hfill
\makebox[0pt]{\hskip#2in
\includegraphics[width=#3\textwidth]{#1}}\vskip#4pt\hfill}


\lefthead{Haiman \& Loeb}
\righthead{Growth of Supermassive Black Holes}

\begin{document}
\title{What is the Highest Plausible Redshift of Luminous Quasars?}
\author{Zolt\'an Haiman\altaffilmark{1}} \affil{Princeton University
Observatory, Princeton, NJ 08544\\ zoltan@astro.princeton.edu}
\vspace{0.5\baselineskip}
\author{Abraham Loeb}
\vspace{0.1\baselineskip}
\affil{Harvard-Smithsonian Center for Astrophysics,
60 Garden St., Cambridge, MA 02138\\
aloeb@cfa.harvard.edu}
\altaffiltext{1}{Hubble Fellow}

\vspace{\baselineskip}
%\submitted{Submitted to ApJ Letters}
\begin{abstract}
The recent discoveries of luminous quasars at high redshifts imply that black
holes more massive than a few billion solar masses have been assembled already
when the universe was less than a billion years old.  We show that the
existence of these black holes is not surprising in popular hierarchical models
of structure formation. For example, the black hole needed to power the quasar
SDSS 1044-0125 at $z=5.8$ can arise naturally from the growth of stellar-mass
seeds forming at $z>10$, when typical values are assumed for the radiative
accretion efficiency ($\sim0.1$), and the bolometric accretion luminosity in
Eddington units ($\sim1$).  Nevertheless, SDSS 1044-0125 yields a non--trivial
constraint on a combination of these parameters. Extrapolating our model to
future surveys, we derive the highest plausible redshift for quasars which are
not lensed or beamed, as a function of their apparent magnitude.  We find that
at a limiting magnitude of $K\sim 20$, quasar surveys can yield strong
constraints on the growth of supermassive black holes out to $z\sim 10$.
\end{abstract}

\section{Introduction}

Quasars have long been believed to be powered by the accretion of gas onto
supermassive black holes (Salpeter 1964; Zel'dovich 1964; Lynden-Bell
1969).  Dynamical studies indicate that remnant black holes (BHs) indeed
reside in the quiescent nuclei of most nearby galaxies (e.g. Magorrian et
al. 1998; Ferrarese \& Merritt 2000; Gebhardt et al. 2000, and references
therein), and imply that BH formation is a generic consequence of galaxy
formation.  Although the activity of bright quasars peaks at $z\sim 2.5$,
quasars are known to exist at higher redshifts.  Approximately 200 quasars
have been found so far at $z>4$ (the majority of which are in the Palomar
Sky Survey\footnote{See http://astro.caltech.edu/\~{}george/z4.qsos for a
comprehensive list of known $z>4$ quasars.}), including the record--holding
bright quasar SDSS 1044-0125 at $z=5.80$, recently discovered by the Sloan
Digital Sky Survey (SDSS; Fan et al. 2000).

Supermassive BHs can grow out of low-mass ``seed'' BHs through accretion or
mergers (Rees 1984; Barkana \& Loeb 2000a).  In popular hierarchical models
for structure formation, the first collapsed gaseous objects have low
masses.  It is natural to identify these galaxies with baryonic masses just
above the cosmological Jeans mass, $\sim 10^{4}~{\rm M_\odot}$, as the
first sites where seed BHs may form (Larson 2000).  Here we postulate that
supermassive BHs are the merger products of individual BHs that grow out of
these seeds through gas accretion.  The natural $e$--folding timescale for
the growth of a single seed can be written as
\begin{equation}
t_{\rm acc}\equiv {M_{\rm bh}\over \dot M_{\rm bh}} = 4 \times 10^7
\left({\epsilon \over 0.1}\right) \eta^{-1}~{\rm yr},
\label{eq:1}
\end{equation}
where $\epsilon\equiv L_{\rm bol}/\dot M_{\rm bh}c^2$ is the radiative
efficiency for a mass accretion rate $\dot M_{\rm bh}$, and $\eta\equiv L_{\rm
bol}/L_{\rm Edd}$ is the bolometric accretion luminosity in Eddington units,
$L_E=4\pi GM_{\rm bh} c \mu_e m_p/\sigma_T$. Here $\sigma_T$ is the Thomson
cross-section and $\mu_e=1.15$ is the mean atomic weight per electron.  The
growth of a $10^9M_\odot$ BH out of a stellar mass seed requires about $\ln
(10^9M_\odot/10M_\odot)=18.4$ $e$--folding times or $\sim 7\times 10^8
(\epsilon/0.1)\eta^{-1}$ yrs. This time is shorter than the age of the universe
at $z=5.8$ if $(\epsilon/0.1)\eta^{-1}\la 1$. However, the time available for
the growth of galaxies is shorter than the entire age of the universe at the
corresponding redshift.

Turner (1991) recognized that exceptionally bright, high--redshift quasars
may yield interesting constraints on cosmic structure formation.  Motivated
by the recent discovery of the SDSS 1044-0125 quasar at $z=5.80$ and the
emergence of a concordance cosmological model for structure
formation\footnote{Throughout this {\it Letter} we adopt a $\Lambda$CDM
cosmology with $\Omega_0=0.3$, $\Omega_\Lambda=0.7$, $\Omega_{\rm
b}=0.045$, $h=0.7$, $\sigma_{8h^{-1}}=0.9$, and $n=1$.}, we revisit this
problem.  In this {\it Letter}, we derive new constraints from the
observed properties of known high-redshift quasars, and assess the
potential for tightening these constraints with future quasar surveys.
%In \S~2, we infer the mass of BHs in observed quasars from their observed
%luminosity, and use a semi-analytic approach to model the BH growth.  The
%resulting constraints on the model parameters are derived in \S~3. Finally,
%\S~4 summarizes our conclusions and their implications.

\section{Cosmological Growth of Quasar Black Holes}

Our constraints derive from the existence of a BH of a particular mass
$M_{\rm bh}$ at redshift $z$, within a galaxy halo of total mass $M_{\rm
halo}$. In the following sub-sections we first express these masses in
terms of observed quantities, and then describe the corresponding
theoretical modeling of the BH growth.

\subsection{Black Hole Mass}

For the high-redshift quasars of interest, we can only infer the BH mass
indirectly from the observed quasar luminosity (although reverberation
mapping provides constraints in other cases; see Kaspi et al. 2000).  Given
the luminosity in the observed band, we first apply a bolometric correction
to obtain the total luminosity $L_{\rm bol}$ (Elvis et al. 1994).  For
example, the approximate spectral slope of $F_\nu \propto \nu^{-1}$ implies
that only $\sim 1\%$ of the bolometric luminosity of SDSS 1044-0125 is in
the SDSS $z^\prime$ band.  The BH mass then follows from the relation,
$L_{\rm bol}=\eta L_{\rm Edd}=1.4\times 10^{38}\eta(M_{\rm bh}/{\rm
M_\odot})~{\rm erg~s^{-1}}$.  In the case of SDSS 1044-0125, we infer
$M_{\rm bh}=3.4 \times 10^9\eta^{-1}{\rm M_\odot}$.  There are no obvious
signs of beaming or lensing in this source. However, if present, both of
these effects would reduce our inferred BH mass or increase our inferred
constraint on $\eta$ by correcting the relation between the apparent and
the intrinsic luminosity of the quasar. In \S 3 we will explore the
sensitivity of our results to such a correction.

\subsection{Black Hole Growth}

In order to model the growth of BHs in our adopted $\Lambda$CDM cosmology, we
rely on the merger history of dark matter halos in the extended
Press--Schechter (EPS) formalism (Press \& Schechter 1974; Bond et al. 1991;
Lacey \& Cole 1993).  We first need to determine the mass of the halo in which
an observed quasar resides.  We estimate $M_{\rm halo}$ based on the halo
abundance within the EPS formalism.  The number of halos with mass $>M_{\rm
halo}$, in a survey of solid angle $\Delta\Omega$, probing a redshift range
$\Delta z$, is given by $N(>M_{\rm halo})=\Delta\Omega\Delta z \times
(dV/d\Omega dz) \times\int_{M_{\rm halo}}^\infty dM [dn/dM(z,M)]$, where
$dV/d\Omega dz$ is the cosmological volume element, and $dn/dM$ is the number
of halos per halo mass per unit volume.  The halo mass for a quasar of
luminosity $L$ is found by equating the observed number of quasars brighter
than $L$ to $f(z)\times N$, where $f(z)$ is the fraction of halos hosting
active quasars.  Because of the exponential shape of the EPS mass function for
rare halos, our inferred halo mass depends only logarithmically on $f(z)$, and
we conservatively adopt $f(z)\sim 1$.  In the case of SDSS 1044-0125, one
bright quasar was found within a $\sim 600$ deg$^2$ survey area at redshift
$z=5.8\pm 0.5$; from this we infer $M_{\rm halo}\approx 1.1\times10^{13}~{\rm
M_\odot}$, with a corresponding velocity dispersion of $\sigma\approx 470~{\rm
km~s^{-1}}$.  It is interesting that the values of $M_{\rm bh}$ and $\sigma$ we
obtain for SDSS 1044-0125 are in good agreement with their relation measured in
local galaxies (Gebhardt et al. 2000; Ferrarese \& Merritt 2000).

We next compute the growth of the central BH mass based on the assembly of
its host halo. Given a halo of total mass $M_{\rm halo}$ at a redshift $z$,
the EPS formalism specifies its average merger history at higher redshifts.
Every branch of this merger tree represents a progenitor of the parent
halo, whose mass is continuously growing through accretion or mergers with
other halos.  To keep our model simple, we assume that every building block
of the original halo develops a seed BH of mass $M_{\rm seed}$ as soon as
it acquires a minimum mass, $M_{\rm min}$, corresponding to a velocity
dispersion $\sigma_{\rm min}$.  The physical motivation for this choice is
that radiative cooling and feedback processes which likely determine
whether or not a massive BH forms in a collapsed halo, depend directly on
$\sigma$ (e.g., Haehnelt et al. 1998).  The mass of each seed BH is assumed
to grow exponentially by accretion, $M_{\rm bh}(t)= \exp[\Delta t(z)/t_{\rm
acc}] M_{\rm seed}$, where $\Delta t$ is the time elapsed between the
formation time of the seed BH and the redshift $z$.  We assume that
eventually all massive BHs merge together to form a single supermassive BH
at the center of the parent halo. As long as the BH mergers are completed
prior to redshift $z$, there is no need to specify when these mergers took
place.  The mass of the resulting BH in the parent halo is the sum of the
individual BHs, each of which has grown by a different amount,
\begin{equation}
M_{\rm bh}(z,M_{\rm halo}) = M_{\rm seed} \int_\infty^z dz^\prime 
\frac{dN_{\rm prog}}{dz^\prime} 
\exp\left[\frac{\Delta t(z,z^\prime)}{t_{\rm acc}}\right],
\label{eq:Mbh}
\end{equation}
where $N_{\rm prog}(z^\prime)$ is the total number of seeded progenitors at
redshift $z^\prime > z$,
\begin{equation}
N_{\rm prog}(z^\prime)=\int_{M_{\rm min}}^{M_{\rm halo}} dM 
\frac{dP(z,z^\prime,M_{\rm halo},M)}{dM}.
\label{eq:Nprog}
\end{equation}
Here $dP(z,z^\prime,M_{\rm halo},M)$ is the number of progenitors in the
mass range between $M$ and $M+dM$ at a redshift $z^\prime$ for a halo whose
mass at redshift $z$ is $M_{\rm halo}$ [see Lacey \& Cole 1993,
Eq. (2.15)]. Note that we adopt the most optimistic assumptions regarding
the BH growth; in reality the growth of each seed BH may be limited by its
fuel reservoir. Also, the central BHs may not get incorporated into a
single supermassive BH during galaxy mergers (Hut \& Rees 1992); however,
for exponential growth, the final BH mass tends to be dominated anyway by
the first seed to have formed.

In summary, our model for the assembly of BHs has four free parameters:
$\sigma_{\rm min}$, $M_{\rm seed}$, $\epsilon$, and $\eta$.  Although the
values of these parameters are a'priori uncertain, their ``natural''
choices are as follows: (i) if the seed BHs are the remnants of massive
stars, then their characteristic mass is $M_{\rm seed}\approx 10~{\rm
M_\odot}$; (ii) a necessary requirement for star--formation is efficient
cooling; in a metal--poor primordial gas with no ${\rm H_2}$ molecules,
this implies a minimum virial temperature of $T\gsim 10^4$K for progenitor
halos with seed BHs, corresponding to a minimum velocity dispersion
$\sigma_{\rm min}\gsim 10~{\rm km~s^{-1}}$; (iii) if the gas accretes to a
non--rotating BH through a steady thin disk, then the radiative efficiency,
$\epsilon\approx 0.06$; and (iv) for high fueling rates bright quasars
would naturally shine close to their limiting luminosity, implying
$\eta\approx 1$.  Variations on these parameter values will be discussed in
\S~3 and \S~4.

\section{Results}

Given the parameters $\sigma_{\rm min}$, $M_{\rm seed}$, $\epsilon$, and
$\eta$, our model yields the mass $M_{\rm bh}$ of the supermassive BH at the
center of a halo of mass $M_{\rm halo}$ at redshift $z$.  In the following, we
derive constraints on these four parameters, by requiring that $M_{\rm bh}$
equals the value estimated for observed high--redshift quasars.

\vspace{\baselineskip} \myputfigure{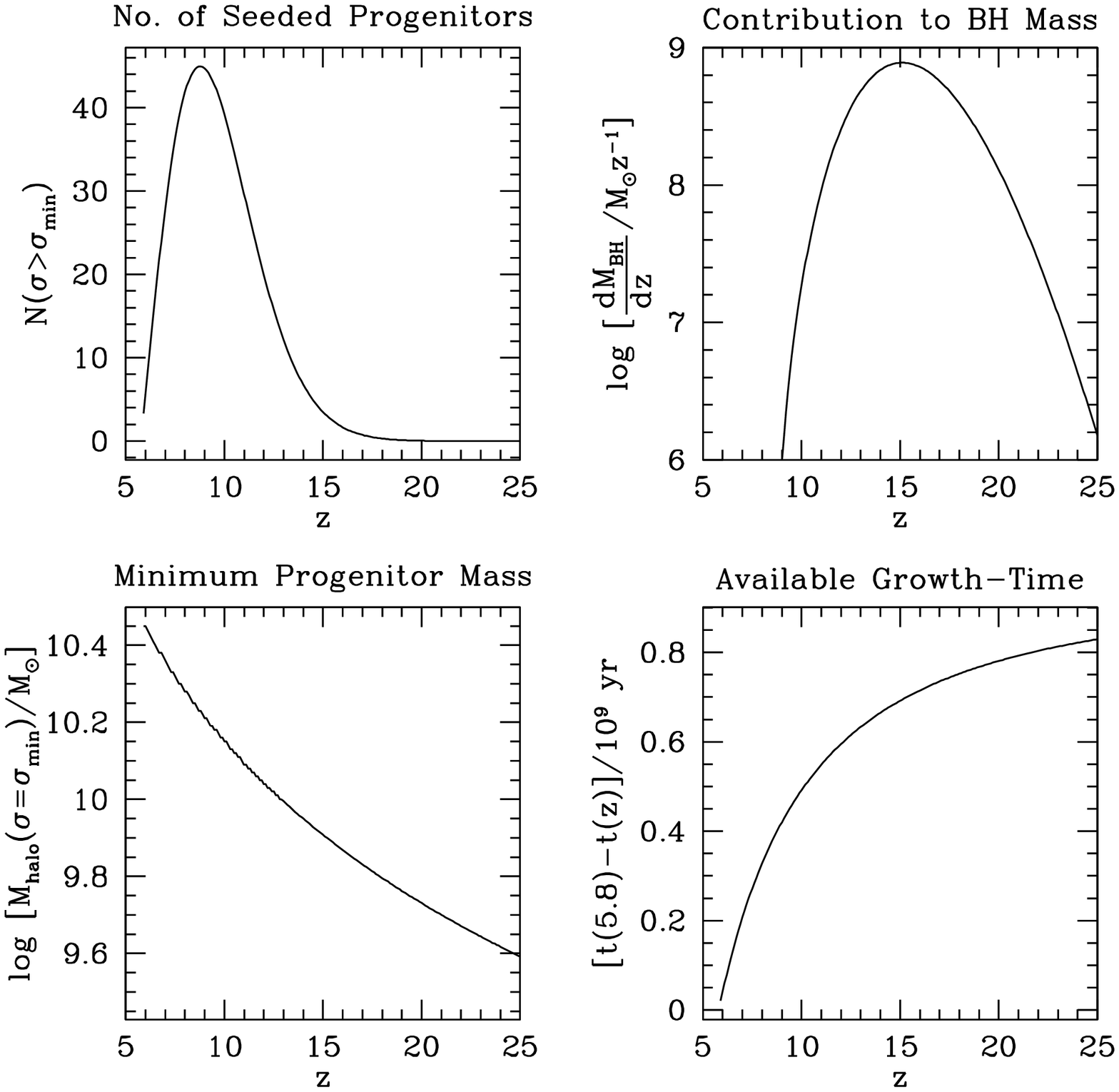}{3.2}{0.45}{-10}{-10}
\figcaption{The assembly history of the black hole (BH) in the SDSS
1044-0125 quasar at $z=5.8$.  The inferred BH mass is $M_{\rm bh}=3.4
\times 10^9\eta^{-1}{\rm M_\odot}$ and the halo mass is
$1.1\times10^{13}~{\rm M_\odot}$. The four panels show, clockwise from top
left, the number of seeded progenitors (those with $\sigma>50~{\rm
km~s^{-1}}$); the contribution from different redshifts to the final BH
mass; the time available for the exponential growth of a seed between $z$
and the redshift of $5.8$; and the halo mass corresponding to $50~{\rm
km~s^{-1}}$ at each redshift.
\label{fig:sdss}}
\vspace{\baselineskip}
 
\subsection{Illustrative Example}

In order to illustrate the BH growth process in our model, we show in
Figure~\ref{fig:sdss} the evolution of various quantities for the SDSS
1044-0125 quasar.  In this example, we have assumed $M_{\rm seed}=10~{\rm
M_\odot}, \epsilon=0.1, \eta=1$, and $\sigma_{\rm min}=50~{\rm km~s^{-1}}$.  As
we find by numerical integration of equations~(\ref{eq:Mbh}) and
(\ref{eq:Nprog}), this combination yields the required BH mass of $M_{\rm
bh}\approx 3.4 \times 10^9~{\rm M_\odot}$ at $z=5.8$.  The top left panel in
Figure~\ref{fig:sdss} shows the number of progenitors of the parent halo
%($M_{\rm halo}\approx 1.1\times10^{13}~{\rm M_\odot}$) 
with a velocity dispersion $>\sigma_{\rm min}$.  For reference, the bottom
left panel shows the corresponding minimum progenitor mass. As the redshift
increases, the number of progenitors increases, peaking at $z\approx 9$,
and then decreasing again as the typical progenitors are broken up into
halos with $\sigma<50~{\rm km~s^{-1}}$.  The top right panel shows the
contribution of progenitors from each redshift to the final BH mass, and
demonstrates that the bulk of the BH mass is contributed by seeds from
$z\approx 15$.  This redshift is considerably higher than the peak at which
most progenitors form.  The increased time available between higher
redshifts and $z=5.8$ (shown explicitly in the bottom right panel) makes
the contribution from the first few progenitors dominant.

\subsection{Constraints from SDSS 1044-0125}

\myputfigure{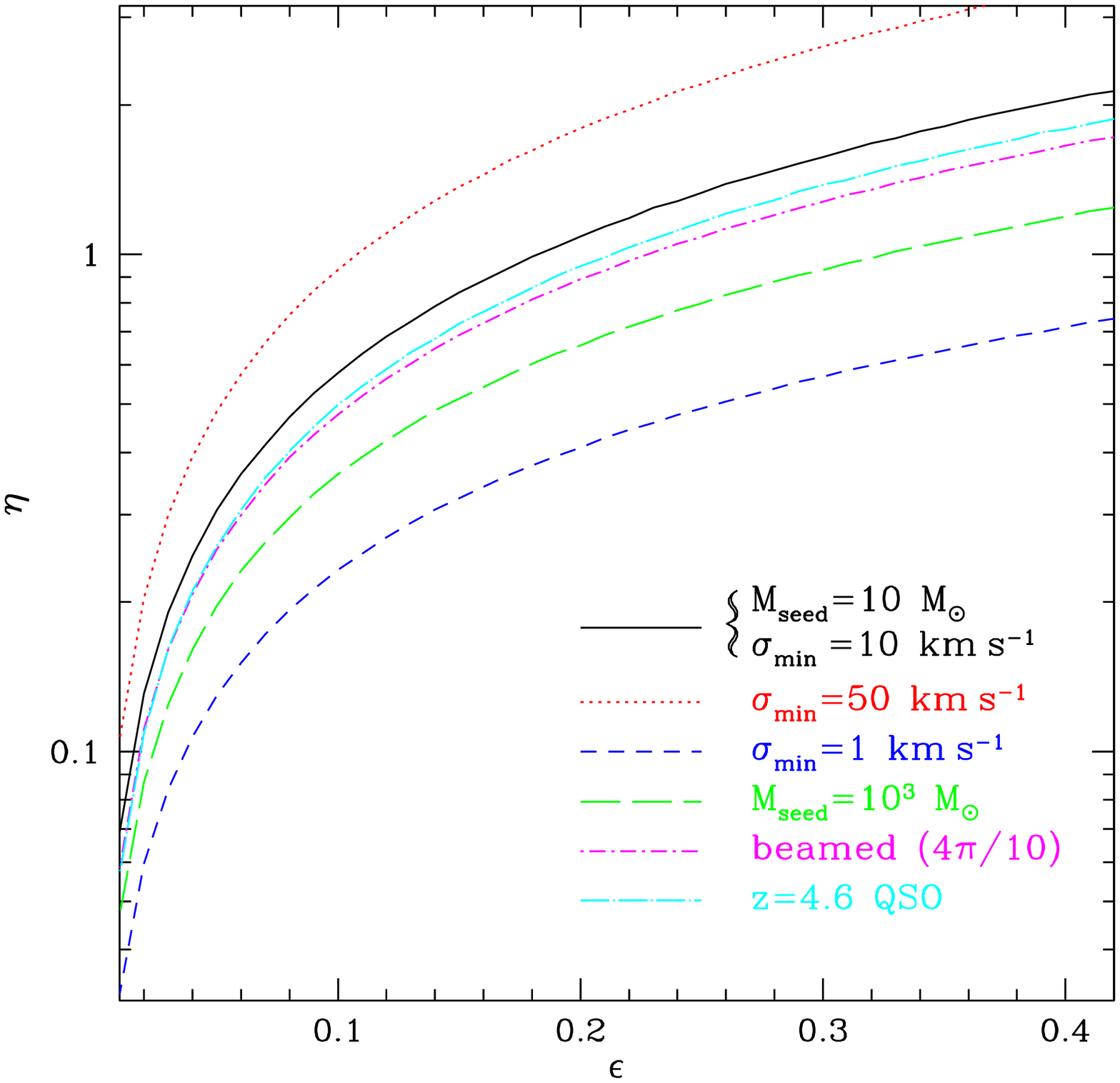}{3.2}{0.45}{-10}{-10} \figcaption{The minimum value
of the typical quasar luminosity in units of the Eddington limit
($\eta\equiv L/L_{\rm Edd}$) as a function of the typical radiative
efficiency of accretion ($\epsilon\equiv L/\dot mc^2$).  The constraints
are derived by requiring that the total BH mass in SDSS 1044-0125 will
build up to its inferred value of $3.4 \times 10^9\eta^{-1}{\rm M_\odot}$
at $z=5.8$.  The solid curve shows a fiducial model with $(M_{\rm seed},
\sigma_{\rm min})=(10~{\rm M_\odot},10~{\rm km~s^{-1}})$; other curves
correspond to variations as labeled.
\label{fig:known}}
\vspace{\baselineskip}

Next, we search for general combinations of $M_{\rm seed}$, $\epsilon$,
$\eta$, and $\sigma_{\rm min}$ that yield the required BH mass for the SDSS
quasar. We solve equation~(\ref{eq:Mbh}) with its left hand side set to
$3.4 \times 10^9\eta^{-1}{\rm M_\odot}$, and the halo mass set to
$1.1\times10^{13}~{\rm M_\odot}$ at $z=5.8$, as discussed above.  For a
given pair of $(M_{\rm seed}, \sigma_{\rm min})$, we then find $\eta$ as a
function of $\epsilon$ by a Newton--Rhapson method.  

The results of this procedure are displayed in Figure~\ref{fig:known} in a
fiducial model with $(M_{\rm seed}, \sigma_{\rm min})=(10~{\rm M_\odot},10~{\rm
km~s^{-1}})$ (solid line) and its variants.  The dotted and short--dashed
curves show results when $\sigma_{\rm min}$ is either increased to $50~{\rm
km~s^{-1}}$ or decreased to $1~{\rm km~s^{-1}}$, respectively.  The high value
applies if the limiting mass for a progenitor halo is determined by feedback
from the UV background following the reionization epoch (Thoul \& Weinberg
1996; Navarro \& Steinmetz 1997; Haiman, Madau \& Loeb 1999); and the low value
applies before reionization if sufficient ${\rm H_2}$ exists to allow cooling
in small halos (Haiman, Abel \& Rees 2000). The long--dashed curve demonstrates
the effect of increasing the typical seed mass to $10^3~{\rm M_\odot}$,
corresponding to remnants of very massive metal-free Population III stars
(VMOs; Bond, Arnett \& Carr 1984; see also Schneider et al. 2000). The
dot--short--dashed curve describes the case where the observed emission from
SDSS 1044-0125 is beamed into a tenth of its sky, so that its true BH mass is
reduced by a factor of 10 and the halo abundance is increased by the same
factor.  Finally, the dot--long--dashed curve shows results for the fiducial
model in the case of the exceptionally bright $z=4.6$ quasar PSS1347+4956.  We
have searched through the list of known $z>4$ quasars and found this object (BH
mass of $5\times 10^{10}~{\rm M_\odot}$, and a host halo mass of $2.7\times
10^{13}~{\rm M_\odot}$ ($\sigma=570~{\rm km~s^{-1}}$, assuming no beaming or
lensing) to provide the second strongest constraint after SDSS 1044-0125.  A
handful of other bright $z>4$ quasars follow closely behind.

Figure~\ref{fig:known} leads to two interesting conclusions.  First, the
set of rather standard values, $M_{\rm seed}=10~{\rm M_\odot}$,
$\sigma_{\rm min}=50~{\rm km~s^{-1}}$, $\epsilon=0.1$, and $\eta=1$, yields
the required BH mass of SDSS 1044-1215.  The corresponding progenitor halos
have cooling times much shorter than the dynamical time, and could
withstand photo-ionization heating by the intergalactic UV background
(e.g. Thoul \& Weinberg 1996; Navarro \& Steinmetz 1997).  In such halos
there is no obvious obstacle to the formation of massive stars, which could
leave behind the required $\sim 10~{\rm M_\odot}$ seed BHs.  Second, the
inferred values of $\eta$ are relatively high, suggesting that BHs must
radiate close to their Eddington limit.  Lower values of $\eta$ are allowed
if $\epsilon\lsim 0.05$; however, such low values of $\epsilon$ would not
account for the total energy output of quasars given the observed mass
density of their remnants (e.g. Ho \& Kormendy 2000; Salucci et al. 1999;
Fabian 2000).  The radiative efficiency may in fact approach a value as
high as $\epsilon=0.42$ for a maximally rotating Kerr BH.

\myputfigure{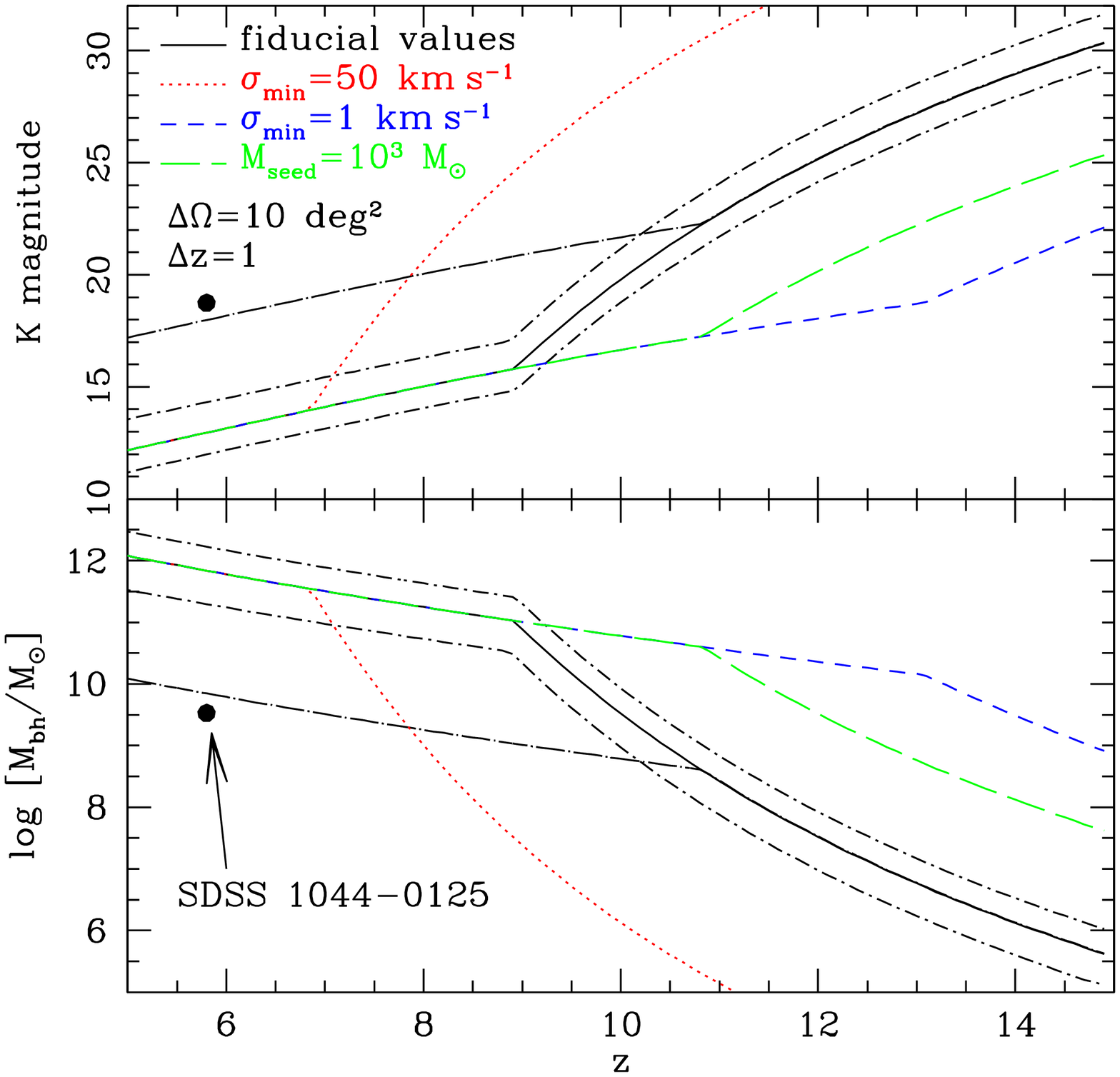}{3.2}{0.45}{-10}{-10} \figcaption{The maximum allowed
BH mass (lower panel), and the corresponding minimum apparent $K$ magnitude
(upper panel), for a single quasar that could be found within a future survey.
The solid curves correspond to our set of fiducial parameter values, $M_{\rm
seed}=10~{\rm M_\odot}$, $\sigma_{\rm min}=10~{\rm km~s^{-1}}$,
$\epsilon=0.06$, and $\eta=1$; the other curves describe variations of this
model as labeled.  The bullets refer to the BH mass and K magnitude inferred
for SDSS 1044-0125.  The survey area is assumed to be 10 deg$^2$, except for
the dot--short--dashed curves which bracket the range between 0.1 (lower curve
on top panel; upper curve in bottom panel) and $10^3$ (upper curve on top
panel; lower curve in bottom panel) deg$^2$.
\label{fig:future}}
\vspace{\baselineskip}

\subsection{Constraints from Future Quasar Surveys}

In anticipation of future results from the SDSS survey as well as from
deeper surveys of high--$z$ quasars, we ask: {\it what is the highest
plausible value of $M_{\rm bh}$ that should be anticipated for a quasar
around a redshift $z$ in a given area on the sky?}  To answer this
question, we compute the mass of the central BH in the most massive dark
halo that could be found in a representative survey area of 10 deg$^2$ and
redshift bin $\Delta z=1$ around different redshifts.  For our set of
fiducial parameter values we take $M_{\rm seed}=10~{\rm M_\odot}$,
$\sigma_{\rm min}=10~{\rm km~s^{-1}}$, $\epsilon=0.06$, and $\eta=1$.  In
Figure~\ref{fig:future}, the solid curve in the lower panel shows the
maximum allowed quasar BH mass in this model as a function of redshift.  In
the upper panel, we show the corresponding apparent $K(AB)$ magnitude with
the corresponding K--correction (Elvis et al. 1994). In our fiducial model,
equation~(\ref{eq:Mbh}) leads to an unphysically large BH mass, in excess
of the observed halo mass.  In Figure~\ref{fig:future} we do not allow the
BH mass to exceed $(\Omega_b/\Omega_0)M_{\rm halo}$.  This additional
constraint appears as the break occurring at $z\approx 9$; BHs below this
redshift are assumed conservatively to have consumed {\it all} the
available gas in their host halo.  Figure~\ref{fig:future} also shows the
maximum allowed BH mass and the corresponding minimum apparent magnitude,
in variants of our fiducial model: examples in which $\sigma_{\rm min}$ is
either increased to $50~{\rm km~s^{-1}}$ or decreased to $1~{\rm
km~s^{-1}}$ (dotted and short--dashed curves), and a model in which the
typical seed mass is increased to $10^3~{\rm M_\odot}$ (long--dashed
curves).  The dot--long--dashed curves show the results when we allow BHs
to consume only 1\% of the available gas in its host halo. Finally, the
dot--short--dashed curves that bracket our fiducial model assume survey
areas of 0.1 deg$^2$ and 1000 deg$^2$.

We find that the survey area enters only logarithmically into our constraints.
However, the allowed BH mass and the resulting apparent magnitude, are strong
functions of redshift.  At its planned sensitivity, the {\it Next Generation
Space Telescope (NGST)} will have a detection threshold of $\sim$32 mag in the
$1-5\mu$m range (with $\lsim 3$ hours of integration and S/N=10).  The assumed
survey size of 0.1 deg$^2$ can be covered by 23 images taken by {\it NGST} in a
total observation time of 2.3 days.  The proposed {\it PRIME} survey\footnote{A
Small Explorer Mission selected by NASA for study. See a description at
http://spacescience.nasa.gov/codesr/smex.} plans to map an area of $\sim 10$
deg$^2$ to $K=27$.  At a limiting magnitude of $K\sim 20~(30)$, future surveys
can yield strong constraints on the growth of supermassive black holes, as they
can map out the ``brightest quasar'' envelope shown in Figure~\ref{fig:future}
out to $z\sim 10~(15)$.

\section{Conclusions}

Existing data on high-redshift quasars implies that BHs as massive as $\sim
3\times10^9~{\rm M_\odot}$ were assembled when the universe was only a tenth of
its present age.  Figure 2 shows that the massive BH inferred for SDSS
1044-0125 at $z=5.8$ can grow in hierarchical galaxy formation models with
plausible parameter values for the initial seed mass ($\sim 10~{\rm M_\odot}$),
the minimum velocity dispersion of collapsed objects that harbor such a seed
($\ga 50~{\rm km~s^{-1}}$), the radiative efficiency ($\sim 6\%$), and the
luminosity in Eddington units ($\sim 1$). Figure 3 illustrates the upper
envelope of plausible luminosity values for high-redshift quasars that might be
found in future surveys.

There are several caveats to the constraints we derived.  First, beaming or
lensing may affect the apparent magnitude of some of the brightest and highest
redshift quasars (Barkana \& Loeb 2000b).  Second, we have assumed that $\eta$
and $\epsilon$ maintain the same values during the luminous quasar phase and
the main growth phase of the BH mass. It is possible, however, that $\epsilon$
is negligibly small during the early growth phase of the BH, and high during
the luminous quasar phase (e.g. Haehnelt \& Rees 1993).  This would be
equivalent to a corresponding increase in the seed BH mass in our model.
Multiple seeds per halo which eventually coalesce are also equivalent to a
single massive seed with the sum of their masses.  Finally, although a recent
cosmological simulation finds good agreement with the Press--Schechter mass
function for the lowest halo masses of interest here at $z=10$ (Jang-Condell \&
Hernquist 2000); the accuracy of this ansatz still remains to be tested over a
wider range of redshifts and halo masses, where there might be systematic
deviations (Jenkins et al. 2000; Sheth \& Tormen 2000).

Recently, various modifications to the standard $\Lambda$CDM model have
been proposed because of a potential conflict between theory and
observations on small spatial scales (e.g., Kamionkowski \& Liddle 2000;
Bode, Ostriker \& Turok 2000; Barkana, Haiman \& Ostriker 2000, and
references therein).  Our constraints are expected to tighten significantly
in such models, since they generically reduce small scale power and thus
eliminate potential hosts of seed BHs at high redshifts. Future surveys of
high-redshift quasars will test these models.

\acknowledgements

We thank Xiaohui Fan, Michael Strauss, and Pat Hall for useful discussions. ZH
was supported by NASA through the Hubble Fellowship grant HF-01119.01-99A,
awarded by the Space Telescope Science Institute, which is operated by the
Association of Universities for Research in Astronomy, Inc., for NASA under
contract NAS 5-26555.  This work was supported in part by NASA grants NAG
5-7039, 5-7768, and NSF grants AST-9900877, AST-0071019 for AL.


\begin{references}

\reference{} Barkana, R., Haiman, Z., \& Ostriker, J. P. 2000, in preparation

\reference{} Barkana, R. \& Loeb, A. 2000a, Physics Reports, in press;
astro-ph/0010468

\reference{} -------------------------. 2000b, ApJ, 531, 613


\reference{} Bode, P., Ostriker, J. P., \& Turok, N. 2000, astro-ph/0010389

\reference{} Bond, J. R., Arnett, W. D., \& Carr, B. J. 1984, ApJ, 280, 825

\reference{} Bond, J. R., Cole, S., Efstathiou, G., Kaiser, N. 1991, 379, 440

\reference{} Elvis, M., et al. 1994, ApJS, 95, 1

\reference{} Fabian, A. C. 2000, in Proc. of X-ray Astronomy'99, Bologna, 1999, in press, astro-ph/0001178

\reference{} Fan, X., et al. 2000, AJ, 120, 1167

\reference{} Ferrarese, L., \& Merritt, D. 2000, ApJL, submitted,
astro-ph/0006053

\reference{} Gebhardt, K. et al. 2000, ApJL, submitted, astro-ph/0006289

\reference{} Haehnelt, M. G., Natarajan, P. \& Rees, M. J. 1998, MNRAS,
300, 817

\reference{} Haehnelt, M. G., \& Rees, M. J.  1993, MNRAS, 263, 168

\reference{} Haiman, Z., Abel, T., \& Rees, M. J. 2000, ApJ, 534, 11

\reference{} Haiman, Z., Madau, P., \& Loeb, A. 1999, ApJ, 514, 535

\reference{} Ho, L. C., \& Kormendy, J. 2000, to appear in The Encyclopedia
of Astronomy and Astrophysics (Institute of Physics Publishing);
astro-ph/0003267

\reference{} Hut, P., \& Rees, M. J.  1992, MNRAS, 259, 27

\reference{} Jang-Condell, H., \& Hernquist, L. 2000, astro-ph/0009254

\reference{} Jenkins, A. et al. 2000, MNRAS, submitted, astro-ph/0005260

\reference{} Kamionkowski, M., \& Liddle, A. R. 2000, Phys. Rev. Lett. 84,
4525

\reference{} Kaspi, S., Smith, P. S., Netzer, H., Maoz, D., Jannuzi, B. T., \&
Giveon, U. 2000, ApJ, 533, 631

\reference{} Lacey, C., \& Cole, S. 1993, MNRAS, 262, 627

\reference{} Larson, R. B. 2000, to be published in the ESA Special
Publications Series (SP-445), edited by F. Favata, A. A. Kaas, and
A. Wilson; astro-ph/9912539

\reference{} Lynden-Bell, D. 1969, Nature, 223, 690

\reference{} Magorrian, J., et al. 1998, AJ, 115, 2285

\reference{} Navarro, J. F., \& Steinmetz, M. 1997, ApJ, 478, 13

\reference{} Press, W. H., \& Schechter, P. L. 1974, ApJ, 187, 425

\reference{} Rees, M. J. 1984, ARA\&A, 22, 471

\reference{} Salpeter, E. E. 1964, ApJ, 140, 796

\reference{} Salucci, P., Szuszkiewicz, E., Monaco, P., \& Danese, L. 1999,
MNRAS, 307, 637

\reference{} Schneider, R., Ferrara, A., Ciardi, B., Ferrari, V., Matarrese, S. 2000, MNRAS, 317, 385

\reference{} Sheth, R. K., \& Tormen, G. 1999, MNRAS, 308, 119 

\reference{} Thoul, A. A., \& Weinberg, D. H. 1996, ApJ, 465, 608

\reference{} Zeldovich, Y. B. 1964, Dok. Akad. Nauk SSSR, 155, 67

\end{references}
\end{document}